\documentclass[twocolumn,prl]{revtex4}
\usepackage{amsmath}
\usepackage{graphicx}

\newcommand{\eq}{\begin{equation}}
\newcommand{\fine}{\end{equation}}

\begin{document}

\title{Experimental test of the no signaling theorem}
\author{Tiziano De Angelis$^1$, Francesco De Martini$^1$, Eleonora Nagali$^1$, and
Fabio Sciarrino $^{2,1}$ \\
$^1$Dipartimento di Fisica dell'Universit\'a ''La Sapienza'' and Consorzio\\
Nazionale Interuniversitario per le Scienze Fisiche della Materia, Roma,
00185 Italy\\
$^2$Centro di\ Studi e Ricerche ''Enrico Fermi'', Via Panisperna
89/A,Compendio del Viminale, Roma 00184, Italy}

\begin{abstract}
In 1981 N. Herbert proposed a gedanken experiment in order to
achieve by the ''First Laser Amplified Superluminal Hookup'' (FLASH)
a faster than light communication (FTL) by quantum nonlocality. The
present work reports the first experimental realization of that
proposal by the optical parametric amplification of a single photon
belonging to an entangled EPR pair into an output field involving
$5\times 10^{3}$ photons. A thorough theoretical and experimental
analysis explains in general and conclusive terms the precise
reasons for the failure of the FLASH program as well as of any
similar FTL proposals.
\end{abstract}

\maketitle

The theory of special relativity lies on a very basic hypothesis:
anything carrying information cannot travel faster than the light in
vacuum. On the other side, quantum physics possesses marked nonlocal
features implied by the Bell's theorem. Nevertheless it has been
shown theoretically, on the basis of several properties of the
Hilbert space that a ''no-signaling theorem'' holds: one cannot
exploit the quantum entanglement between two space-like separated
particles in order to realize any ''faster-than-light'' (FTL)
communication \cite{Ghir83,Bru00,Peres83}. Indeed the quantum theory
of communications teaches us that a single particle cannot carry
information about its coding basis at the receiving station. Then,
in an attempt to break the ''peaceful coexistence'' between quantum
mechanics and relativity, Nick Herbert proposed in 1981 a feasible
experimental scheme, i.e. a FLASH machine, the acronym standing for
''First Laser-Amplified Superluminal Hookup'' \cite{Herb02}. The
proposal, based on the amplification by stimulated emission of a
photon in an entangled state, elicited a debate among theorists
leading at last to the formulation of the quantum ''no-cloning
theorem'' \cite{Woot82,Pere03}. However, recent studies have shown
that the amplifier noise related to the reduction of''fidelity''
$F<1$ implied by the ''no-cloning theorem'' cannot be held
responsible for the failure of the FLASH program, since the
realizable ''optimal'' fidelity can be large enough to provide a
sufficiently large signal-to-noise ratio of the output signals \cite
{DeMa05,Scar05,Cerf06}. In spite of that, due to the complex and
subtle dynamics underlying any real amplification process, a clear
and unambiguous theoretical argument on the precise cause of that
failure has never been singled out. After so many years a recourse
to the experiment and a careful theoretical analysis were held
necessary. This will contribute to our understanding on the
properties of the distribution of quantum information within a more
advanced theory of the cloning process.

\begin{figure}[h]
\includegraphics[scale=.49]{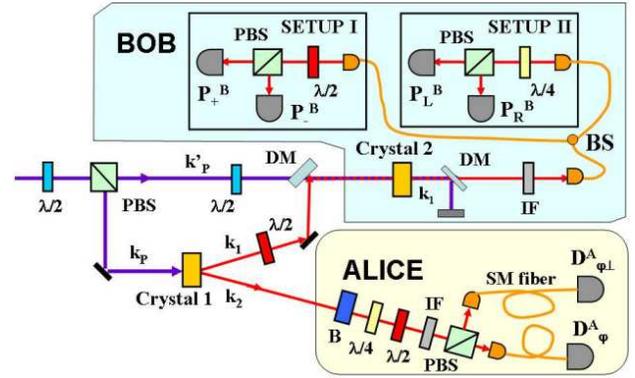}
\caption{Configuration of the quantum injected optical parametric amplifier.
The polarization entangled photon couple is generated by crystal 1. Crystal
2, realizing the OPA action, is cut for collinear type II phase matching. }
\label{CalibrazFoto}
\end{figure}

The setup proposed by Herbert is reported in Fig.1. Two space-like
distant observers, say Alice and Bob share two polarization
entangled photons generated by a common EPR source (Crystal 1).
Alice detects by the
phototubes $D_{\varphi }^{A}$ and $D_{\varphi \bot }^{A}$ the polarization $%
\overrightarrow{\pi }_{A}$ of her photon in any two orthogonal
measurement basis $\left\{ \overrightarrow{\pi },\overrightarrow{\pi }%
^{\perp }\right\} _{A}$. Let us refer for simplicity and with no lack of
generalization to the following states:$\;$linear \textit{polarizations%
} $\{\overrightarrow{\pi }_{\pm }=2^{-1/2}(\overrightarrow{\pi }_{V}\pm
\overrightarrow{\pi }_{H})\}_{A}$ and \textit{circular} $\overrightarrow{\pi
}^{\prime }s$ $\{\overrightarrow{\pi }_{R}$=$2^{-1/2}(\overrightarrow{\pi }%
_{H}+i\overrightarrow{\pi }_{V})$, \ $\overrightarrow{\pi }_{L}$=$2^{-1/2}(%
\overrightarrow{\pi }_{V}-i\overrightarrow{\pi }_{H})\}_{A}$, where $%
\overrightarrow{\pi }_{H}$ and $\overrightarrow{\pi }_{V}$ are horizontal
and vertical $\overrightarrow{\pi }^{\prime }s$, respectively. Consider that
the choice of the basis is the only coding method accessible to Alice in
order to establish a meaningful communication with Bob. If Bob could guess
the coding basis chosen by Alice, then a FTL signaling process would be
established. Since the detection of an unknown \textit{single} particle
cannot carry any information on the coding basis \cite{Peres83}, Herbert
proposed that Bob could make a ''new kind'' of measurement on the photon
through the \textit{amplification }by a ''nonselective laser gain tube''. In
the jargon of modern quantum optics this consists of a \textit{universal,}
i.e. polarization independent amplifier. The amplified beam is then split by
a symmetric beam-splitter (BS)\ so that Bob can perform a measurement on
half of the amplified particles in the $\left\{ \overrightarrow{\pi }_{\pm
}\right\} _{B}$ basis, and on the other half in the $\left\{ \overrightarrow{%
\pi }_{R},\overrightarrow{\pi }_{L}\right\} _{B}$ basis. The couples of
electronic signals registered by two couples of photodetectors are $\left\{
I_{+}^{B},I_{-}^{B}\right\} \;$and $\left\{ I_{L}^{B},I_{R}^{B}\right\} $,
correspondingly. In order to challenge the Herbert's proposal, physicists
have resorted to the \textit{no-cloning theorem }which forbids the perfect
duplication of quantum states. Actually Herbert, aware of the impossibility
to produce perfect clones of any input qubit because of the noisy
contributions from spontaneous emission, proposed to extract \textit{%
appropriate signal signature \ }implying a\textit{\ }discrimination between
the two measured quantities $\langle \left| I_{+}^{B}-I_{-}^{B}\right|
\rangle $ and $\langle \left| I_{R}^{B}-I_{L}^{B}\right| \rangle $ which
should depend on the coding basis chosen by Alice.

Let's account now for the experiment. We first selected a definite model for
the amplifier by which an experimental test should be carried out. A laser
beam at wavelength (wl) $\lambda _{P}=397.5nm$ was split in two beams
through a $\lambda /2$ waveplate and a polarizing beam splitter (PBS) and
excited two non-linear (NL) crystals ($\beta $-barium borate) cut for type
II phase-matching \cite{DeMa05}. Crystal 1, excited by the beam $\mathbf{k}%
_{P}$, is the spontaneous parametric down-conversion source of entangled
photon couples of wl $\lambda =2\lambda _{P}$, emitted over the two output
modes $\mathbf{k}_{i}$ ($i=1,2$) in the \textit{singlet} state $\left| \Psi
^{-}\right\rangle _{k1,k2}$=$2^{-%
{\frac12}%
}\left( \left| H\right\rangle _{k1}\left| V\right\rangle _{k2}-\left|
V\right\rangle _{k1}\left| H\right\rangle _{k2}\right) \label{SPDCentangled}$%
. The single photon state generated over the mode $\mathbf{k}_{1}$ was
directed toward Bob's station, simultaneously with a pump beam (mode $%
\mathbf{k}_{P}^{\prime }$). In virtue of the nonlocal correlation acting on
modes $\mathbf{k}_{1}$ and $\mathbf{k}_{2}$, the input qubit on mode $%
\mathbf{k}_{1}$ was codified in the polarization state $\overrightarrow{\pi }
$ by measuring the photon on mode $\mathbf{k}_{2}$ in the appropriate
polarization basis. The injected photon and the UV pump beam $\mathbf{k}%
_{P}^{\prime }$ were superposed by a dichroic mirror ($DM$) with high
reflectivity at $\lambda$ and high transmittivity at $\lambda _{P} $.

Let us now describe how the optical amplifier, i.e. the
\textit{cloning machine}, has been realized \cite{DeMa05}. We
adopted a quantum injected - optical parametric amplifier (QI-OPA),
exploiting the process of stimulated emission in a NL crystal
(crystal 2) in the highly efficient collinear regime, generating a
large number of photons over the same
direction \cite{DeMa98B}. This transformation, often referred to as \textit{%
phase-covariant quantum cloning }(PCQC) , realizes cloning with
constant amplification for qubits belonging to the equatorial plane
orthogonal to the z-axis of the corresponding Bloch sphere
\cite{Brus00}. Since they are identified only by a phase $\varphi $,
the qubits belonging to this plane will be also referred to as:
$\left| \varphi \right\rangle =2^{-1/2}(\left|
H\right\rangle +e^{i\varphi }\left| V\right\rangle )$. Note that both the $%
\left\{ \overrightarrow{\pi }_{\pm }\right\} $and $\left\{ \overrightarrow{%
\pi }_{R},\overrightarrow{\pi }_{L}\right\} $ polarization bases do
belong to that plane. The adoption of the \textit{optimal} PCQC
scheme exhibits the highest quality of the output clones allowed by
quantum theory \cite{Scia05}.

The interaction Hamiltonian accounting for the parametric amplification of
crystal 2, reads $\widehat{H}=i\chi \hbar \left( \widehat{a}_{H}^{\dagger }%
\widehat{a}_{V}^{\dagger }\right) +h.c.$ and acts on the single spatial mode
$\mathbf{k}_{1}$ where $\widehat{a}_{\pi }^{\dagger }$ is the one photon
creation operator associated to mode $\mathbf{k}_{1}$ with polarization $%
\overrightarrow{\pi }$ \cite{DeMa98B}. Its form implies the phase-covariance
and the optimality of the cloning process. The multi-photon output state $%
\left| \Phi ^{\varphi }\right\rangle $ generated by the OPA amplification
injected by a single photon qubit $|\varphi \rangle $ is found $\left| \Phi
^{\varphi }\right\rangle =\widehat{\mathbf{U}}\left| \varphi \right\rangle$
where $\widehat{\mathbf{U}}=\exp (-i\widehat{H}\hbar ^{-1}t)$\ is the
unitary transformation implied by the QIOPA\ process \cite{DeMa98B}.
Consider any two generic qubits $\left| \varphi \right\rangle \;$and $\left|
\psi \right\rangle \;$on the Alice's Bloch sphere. They are connected by the
general linear transformation: $\left| \psi \right\rangle \;$= $(\alpha
\left| \varphi \right\rangle +\beta \left| \varphi \bot \right\rangle )\;$%
with\ $\left| \alpha \right| ^{2}+\left| \beta \right| ^{2}$=$1$. The
orthogonal qubits are obtained by the \textit{anti-unitary} time-reversal
transformation: $\left| \varphi ^{\bot }\right\rangle \;$=\textbf{T}$\left|
\varphi \right\rangle $ and $\left| \psi ^{\bot }\right\rangle \;$=\ \textbf{%
T}$\left| \psi \right\rangle =$ $(-\beta ^{\ast }\left| \varphi
\right\rangle +\alpha \ast \left| \varphi \bot \right\rangle )$. In
addition, the \textit{unitarity} of $\widehat{\mathbf{U}}$ \ leads to: $%
\left\langle \Phi ^{\varphi }\mid \Phi ^{\varphi }\right\rangle \;$=$\;1$, $%
\left\langle \Phi ^{\varphi \bot }\mid \Phi ^{\varphi }\right\rangle
=0$:

\begin{equation}
\begin{aligned}
\left| \Phi ^{\psi }\right\rangle \text{\ =\ }(\alpha \left| \Phi
^{\varphi
}\right\rangle +\beta \left| \Phi ^{\varphi \bot }\right\rangle )\text{;}%
\
\\
\left| \Phi ^{\psi \bot }\right\rangle \;\text{= \textbf{T}}\left|
\Phi ^{\psi }\right\rangle \;\text{=\ }(-\beta ^{\ast }\left| \Phi
^{\varphi
}\right\rangle +\alpha ^{\ast }\left| \Phi ^{\varphi \bot }\right\rangle )%
\text{ }
\end{aligned}
\end{equation}
Note that the multi-particle state $\left| \Phi ^{\varphi
}\right\rangle $, sometimes dubbed ''massive qubit'', bears exactly
the same quantum dynamical properties of the injected
single-particle qubit $\left| \varphi \right\rangle $ \cite{DeMa06}.
The average photon number $M_{\pm }^{B}$ over $\mathbf{k}_{1}$ with
polarization $\overrightarrow{\pi }_{\pm }$ is found to depend from
the phase $\varphi $ of the input qubit as follows: $M_{\pm
}^{B}(\varphi )\;$= $\overline{m}+\frac{1}{2}(2\overline{m}+1)(1\pm
\cos \varphi )$ with $\overline{m}=\sinh ^{2}g$, $g$ being the NL
gain of the OPA. In the analysis basis of circular polarization the
expression of the average photon number $M_{R,L}^{B}$ is the same
except for a $\frac{\pi }{2}$ phase $\varphi $-shift.

The output state of the crystal $2$ with wl $\lambda $ was spatially
separated by the fundamental UV beam through a dichroic mirror, spectrally
filtered by an interferential filter ($IF$) with bandwidth $1.5nm,$ coupled
to a single mode fiber (SM) and split over two output modes by a
beam-splitter (BS). Then each output field was polarization analyzed and
detected by a couple of photomultiplier (PM) tubes ($P_{+/R}^{B}$ and $%
P_{-/L}^{B}$). The adopted PM's were Burle C31034-A02 with Ga-As
cathode and quantum efficiency $\eta _{QE}=13\%$. The pulse signals
were registered by a digital memory oscilloscope (Tektronix
TDS5104B) triggered by $\{D_{\varphi }^{A},D_{\varphi \bot }^{A}\}$.

\begin{figure}[h]
\includegraphics[scale=.18]{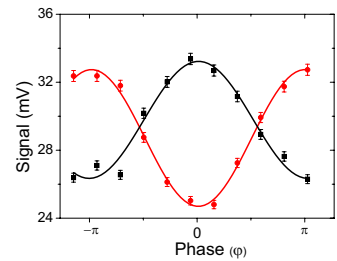}
\caption{Average signal versus the phase of the input qubit: $I_{+}^{B}$
(square marks), $I_{-}^{B}$ (circle marks). }
\label{CalibrazFoto}
\end{figure}

We estimated the experimental gain value $g=\left( 4.45\pm 0.04\right) $,
which corresponds to a generated mean photon number about equal to $4000$ in
the stimulated regime. The coherence property of the multiphoton output
field $\left| \Phi ^{\varphi }\right\rangle $ implied by the quantum
superposition character of the input qubit $\left| \varphi \right\rangle $
was measured in the basis $\overrightarrow{\pi }_{\pm }$ over the output
''cloning'' mode $\mathbf{k}_{1}$ by signals registered by $P_{+}^{B}$ and $%
P_{-}^{B}$ \textit{conditionally} to single photon detection event by $%
D_{\varphi \bot }^{A}$. The corresponding signals $I_{+}^{B}$ and
$I_{-}^{B}$, averaged over $2500$ trigger pulses, are reported
versus the phase $\varphi $ in Fig.2. The typical fidelity values of
the output photon polarization state have been found to be $0.58\div
0.60$ within the theoretical figures. The discrepancy between
experimental and theoretical
values can be attributed to the reduced \textit{visibility} of input qubit ($%
\mathcal{V}_{in}\approx 85\%$) and to the reduced probability $p\simeq 0.4$
of correct amplifier injection under $D_{\varphi }^{A}$-trigger detection. A
theoretical model including these imperfections gives rise to an expected
value for the visibility $\widetilde{\mathcal{V}}=\mathcal{V}_{in}\ [p(2%
\overline{m}+1)]/[p(2\overline{m}+1)+2\overline{m}]$ which fits the
experimental data obtained for different values of $g$.

As first step toward the FLASH test, we analyzed the correlations
between Alice and Bob's measurements by a \textit{conditional
}experiment. Let's make the hypothesis that Bob analyzes his field
in the basis $\{\overrightarrow{\pi }_{\pm }\}_{B}$, by the
experimental Setup I in Fig.1. By this one the observable $\Delta
_{+,-}^{B}$=$(I_{+}^{B}-I_{-}^{B})$ was measured over 2500 equally
prepared
experiments, for different input qubit states $\left| \varphi \right\rangle $%
. As shown by the interference pattern given in Fig.3-\textbf{a} the average
difference signal $\left\langle \Delta _{+,-}^{B}\right\rangle $ is equal to
zero for an input state belonging to the basis $\left\{ \overrightarrow{\pi }%
_{R},\overrightarrow{\pi }_{L}\right\}$, i.e. for $\varphi =-\frac{\pi }{2},%
\frac{\pi }{2}$, while it achieves well defined maximum and minimum values
for input qubits equal to $\left| +\right\rangle $ $(\varphi =0)$ or $\left|
-\right\rangle $ $(\varphi =\pi )$, respectively. According to Herbert's
proposal, the average of the \textit{moduli} of $\Delta _{+,-}^{B}$ , i.e.
the value of $\left\langle |\Delta _{+,-}^{B}|\right\rangle $ has been
estimated. More precisely, in order to further reduce the effects of the
shot-to-shot amplification fluctuations, we registered the average value $%
\langle |N_{+,-}|\rangle $ with $%
N_{i,j}=(I_{i}^{B}-I_{j}^{B})/(I_{i}^{B}+I_{j}^{B})$ for different values of
the phase of the input qubit. As shown by Fig.3-\textbf{b}, even in a
\textit{conditional} experiment any information on the input state $\left|
\varphi \right\rangle $ is deleted by the averaging process then making all
different input $\left| \varphi \right\rangle $ states fully
indistinguishable. By a similar procedure, in correspondence with the $\{%
\overrightarrow{\pi }_{R},\overrightarrow{\pi }_{L}\}_{B}$ basis we observed
that no information on the basis could be drawn by $\left\langle
|N_{R,L}|\right\rangle $.

\begin{figure}[t]
\includegraphics[scale=.33]{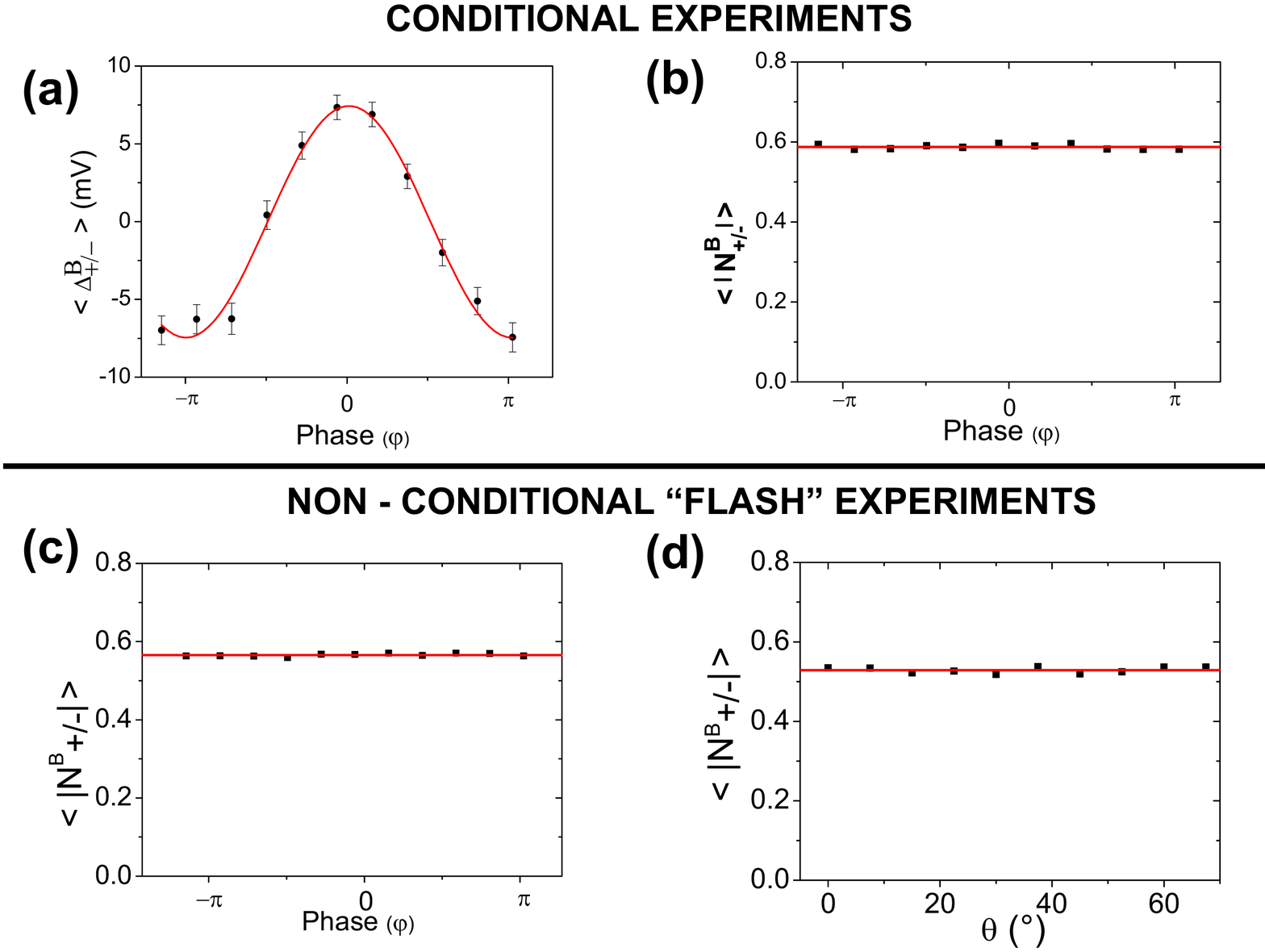}
\caption{ (\textbf{a}) Mean value of $\Delta _{+,-}^{B}$ for different input
qubit states. (\textbf{b}) Mean value of $|N_{+,-}|$ versus the phase of the
input qubit. Error bars are within the corresponding experimental points. (%
\textbf{c}) Mean value of $|N_{\pm }|$ versus the Alice measurement basis $\{%
\vec{\protect\pi}_{\protect\varphi },\vec{\protect\pi}_{\protect\varphi \bot
}\}_{A}$ . (\textbf{d})Mean value of $|N_{\pm }|$ versus the Alice
measurement basis: $\{\vec{\protect\pi}_{\protect\theta },\vec{\protect\pi}_{%
\protect\theta \bot }\}_{A}$.}
\label{CalibrazFoto}
\end{figure}

\begin{figure}[t]
\includegraphics[scale=.3]{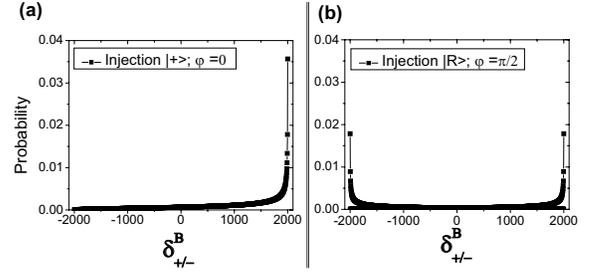}
\caption{Theoretical probability distribution of $\protect\delta^{B}_{+/-}$
for a fixed number of overall output clones $n_{+}^{B}+n_{-}^{B}$equal to
2001. (a) Injected state: $|+\rangle$. (b) Injected state: $|R\rangle$.}
\label{fig4}
\end{figure}

In order to understand the previous results, we investigate the dynamics of
the process by considering the probability distribution $\mathcal{P}_{\pm
}^{\varphi }(x)$ of the fluctuating observable $x\equiv \delta _{\pm }^{B}$
= $n_{+}^{B}-n_{-}^{B}$ for an input qubit with phase $\varphi $. There $%
n_{\pm }^{B}$ is the number of photons with polarization $\overrightarrow{%
\pi }_{\pm }$. Assume the \textit{polarization} basis
$\{\overrightarrow{\pi }_{\pm }\}_{A}$ \ corresponding to the
\textit{phase} basis $\{\varphi =0,\varphi ^{\bot }=\pi \}_{A}$. As
shown by Fig.4-a, the probability distribution $\mathcal{P}_{\pm
}^{0}(x)$ exhibits a peak in correspondence of $\widetilde{x}^{0}>0$
while the distribution$\;\mathcal{P}_{\pm }^{\pi }(x)$ (not reported
in the Figure) is identical to $\mathcal{P}_{\pm
}^{0}(x)$ but reversed respect to the axis $x=0$ with a peak at $\widetilde{x%
}^{\pi }=-\widetilde{x}^{0}$. The ensemble average values of the observable $%
x$ are $\left\langle x\right\rangle ^{0}=2\overline{m}+1$ and $\left\langle
x\right\rangle ^{\pi }=-(2\overline{m}+1)$, respectively. Suppose now that
Bob keeps his previous measurement basis $\{\overrightarrow{\pi }_{\pm
}\}_{B}$ but Alice adopts the new coding basis $\{%
\overrightarrow{\pi }_{L/R}\}_{A}$ $\Rightarrow $ $\{\varphi =\frac{\pi }{2}%
,\varphi ^{\bot }=-\frac{\pi }{2}\}_{A}$. The average values read now $%
\left\langle x\right\rangle ^{\frac{\pi }{2}}=\left\langle x\right\rangle ^{-%
\frac{\pi }{2}}=0$ and the corresponding distributions $\mathcal{P}_{\pm
}^{\pm \frac{\pi }{2}}(x)$ are equal for both phases $\pm \frac{\pi }{2}$.
Furthermore, Fig. 4-b shows that the two $\mathcal{P}_{\pm }^{\pm \frac{%
\pi }{2}}(x)$ do not exhibit a single peak centered around the common
average value $\widetilde{x}^{\pm \frac{\pi }{2}}=0$, as expected for any
gaussian-like distribution, but two symmetrical and balanced peaks that are
in exact correspondence with the ones shown for $\widetilde{x}^{0}$ and $%
\widetilde{x}^{\pi }$, i.e. in correspondence with the former Alice
polarization basis $\{\overrightarrow{\pi }_{\pm }\}_{A}$. Let's now
focus our attention on the probability distribution of $\left|
\delta _{\pm }^{B}\right| $ pointing out its implication on FTL
communication. From the peculiar behavior of the
$\mathcal{P}_{\pm}^{\varphi }(x)$ shown in Fig.4, it is
straightforward to conclude that any input qubit leads to the same
distribution for $\left| \delta _{\pm }^{B}\right| $. Indeed this
unexpected and somewhat counterintuitive result, confirmed by
detailed calculations, lies at the basis of the explanation for the
FLASH\ failure. We consider for simplicity and without loss of
generality, the average of the square function of $\delta _{\pm
}^{B}$, which is equivalent for our purposes to the average of $\
\left| \delta _{\pm }^{B}\right| $. It
is found $\langle (\delta _{\pm }^{B})^{2}\rangle _{\varphi }=12\overline{m}%
^{2}+12\overline{m}+1$ for any input qubit $|\varphi \rangle $
injected on mode $\mathbf{k}_{1}$ pointing out the phase
independence of the function. Let us note that the previous results
are not invalidate by the low detection efficiency $\eta $. Indeed
the observable $\langle (\delta _{\pm }^{B})^{2}\rangle _{\varphi }$
behaves as $\overline{m}^{2}$, corresponding to a super-poissonian
statistics of $\delta _{\pm }^{B}$. At variance with the squeezing
phenomenology, its fluctuation properties can be investigated with
$\eta <<1$.

Let's now account for the set of experiments suggested by the FLASH
proposal. In this case, since no classical information channel
should connect Alice to Bob, a \textit{non-conditional} experimental
configuration is necessary. This test was realized by adopting two
different triggering solutions for the registering setups at Bob's
site: (1) The trigger was provided by the output of a XOR gate fed
by the output TTL signals registered by $\{D_{\varphi }^{A}$ and
$D_{\varphi \bot }^{A}\}$. This device ensured a good discrimination
against noise, but any information about Alice's basis was erased.
(2) Any channel of information between Alice and Bob was accurately
severed. Apart from noise, the two solutions led to identical
results. Two experiments were realized. (a) The average value of
$|N_{\pm }|$ was measured with Alice measuring her photon
polarization in any basis $\{\overrightarrow{\pi }_{\varphi },%
\overrightarrow{\pi }_{\varphi \bot }\}_{A}$ with $\overrightarrow{\pi }%
_{\varphi }=2^{-1/2}(\overrightarrow{\pi }_{H}+e^{i\varphi }\overrightarrow{%
\pi }_{V})$: Fig. 3-c. (b) The value of $\left\langle |N_{\pm
}|\right\rangle $ was measured with Alice measuring her photon polarization
in any basis $\{\overrightarrow{\pi }_{\vartheta },\overrightarrow{\pi }%
_{\vartheta \bot }\}_{A}$ belonging to the equatorial plane
orthogonal to the y-axis. There: $\overrightarrow{\pi }_{\theta
}=(\cos \theta \overrightarrow{\pi }_{H}+\sin \theta
\overrightarrow{\pi }_{V})$: Fig.3-d. In both cases the observable
$|N_{\pm }|$ was found to be independent from Alice's choice of
measurement basis.

 All these results may be understood in very general and conclusive
terms on the basis of the fundamental \textit{linearity} of Quantum
Mechanics. Suppose that the Alice's polarization coding bases
consists of any generic
couple of qubit sets: $\left\{ \overrightarrow{\varphi },\overrightarrow{\varphi }%
\bot \right\} _{A}$and $\left\{ \overrightarrow{\chi },\overrightarrow{\chi }%
\bot \right\} _{A}$. In any \textit{non conditional} experiment Bob must
carry out his measurements on the state he receives, which here are either
one of the two \textit{mixed} mesoscopic states: $\rho _{B}$ = $%
{\frac12}%
(\left| \Phi ^{\varphi }\right\rangle \left\langle \Phi ^{\varphi
}\right| $+$\left| \Phi ^{\varphi \bot }\right\rangle \left\langle
\Phi ^{\varphi \bot }\right| )$ \ or
$\rho _{B^{\prime }}$ =$\;%
{\frac12}%
(\left| \Phi ^{\chi }\right\rangle \left\langle \Phi ^{\chi }\right| $+$%
\left| \Phi ^{\chi \bot }\right\rangle \left\langle \Phi ^{\chi \bot
}\right| )$. However, on the basis of Equation 1 is found: $\rho
_{B}=\rho _{B^{\prime }}=\hat{1}$, i.e. a fully mixed states.  In
other words, referring to previous considerations,  the linearity of
the Hilbert space requires that the \textit{sum} of the
\textit{probability distributions} $\mathcal{P}^{(\varphi , \varphi
\bot)}_{\pm}(x)=\mathcal{P}^{\varphi}_{\pm}(x)+ \mathcal{P}^{\varphi
\bot}_{\pm}(x)$ of the variable $x$ is invariant respect to the
corresponding coding basis $\left\{ \varphi ,\varphi \bot \right\}
_{A}$. Since this result is valid for any
measurement setup chosen by Bob, the discrimination of the phase $%
\varphi $ is impossible, in spite of the different mean values
$\langle x\rangle^{\varphi}$ found for different $\varphi ^{\prime
}s$ in the fringing patterns of Fig.3a. This result, shown by the
experimental data of Fig. 3c-d, is general and prevents any
superluminal communication based on quantum nonlocality.

At last, let's consider again the role of the ''no cloning
theorem'', itself a consequence of the linearity of quantum
mechanics. Indeed the limitations implied by the cloning dynamics
are not restricted to the bounds on the cloning ''fidelity'', as
commonly taken from granted. They also largely affect the
\textit{high-order correlations} among the different clones. In
facts noisy but separable copies would lead to a perfect state
estimation for $g\rightarrow \infty $. However the particles
produced by a cloning machine are far from being mutually
independent and the corresponding states are the \textit{worst} ones
in view of estimating the reduced density matrix, as shown by
\cite{Demk05}. The correlations among the different clones, a
consequence of the macroscopic coherence of the output field, are
clearly exhibited by the peculiar distribution functions of Fig.4.
This interpretation has been confirmed in our laboratory by a side
quantum state-tomography experiment based on the technique
introduced in \cite{Cami06}, showing that the density matrix of a
two clones system $\rho _{C}^{I,II}$ is clearly different from the
case of separable clones: $\rho _{C}^{I}\otimes \rho _{C}^{II}$.

In summary, we have presented a conclusive theoretical and
experimental investigation of the physical processes underlying the
failure of the FLASH program. Since the relevant quantum
informational aspects of the correlations implied by the cloning
process have been little explored in the literature, the present
work may elicit an exciting new trend for future research. We thank
Gerd Leuchs, Wojciech Zurek, Giancarlo Ghirardi, and Hans Briegel
for interesting discussions. We also thank Chiara Vitelli, Sandro
Giacomini and Giorgio Milani for technical support. We acknowledge
support from the MIUR (PRIN 2005) and from CNISM(Progetto Innesco
2006).

\end{document}